\documentclass[final]{cvpr}

\usepackage{times}
\usepackage{epsfig}
\usepackage{graphicx}
\usepackage{amsmath}
\usepackage{amssymb}
\usepackage{subfigure}
\usepackage{overpic}

\usepackage{bm}
\usepackage{algorithm}

\usepackage{algorithmicx}
\usepackage{algpseudocode}

\usepackage{enumitem}
\setenumerate[1]{itemsep=0pt,partopsep=0pt,parsep=\parskip,topsep=5pt}
\setitemize[1]{itemsep=0pt,partopsep=0pt,parsep=\parskip,topsep=5pt}
\setdescription{itemsep=0pt,partopsep=0pt,parsep=\parskip,topsep=5pt}

\usepackage[pagebackref=true,breaklinks=true,colorlinks,bookmarks=false]{hyperref}

\def\cvprPaperID{159} 
\def\confYear{CVPR 2020}

\graphicspath{{figures/}}

\newcommand{\Ignore}[1]{}

\newcommand{\NullFigure}[3]{}

\title{Rule-based Procedural Tree Modeling Approach}

\author{papers 0578}
\author{Yinhui Yang ~~~~~ Rui Wang ~~~~~ Yuchi Huo
}

\begin{document}

\maketitle

\begin{abstract}
In some entertainment and virtual reality applications, it is necessary to model and draw the real world realistically, so as to improve the fidelity of natural scenes and make users have a better sense of immersion. However, due to the morphological structure of trees The complexity and variety present many challenges for photorealistic modeling and rendering of trees. This paper reviews the progress achieved in photorealistic modeling and rendering of tree branches, leaves, and bark over the past few decades. The main achievement is mainly a rule-based procedural tree modeling method.
\end{abstract}

\section{Introduction}

In recent years, with the vigorous development and popularization of industries such as virtual reality and CG movies, people's demand for rich digital content is increasing day by day. In some virtual reality applications related to real scenes, people's requirements for model realism are also increasing. High. As an important part of the real world, the realistic modeling and presentation of trees will greatly enhance the sense of reality and immersion in the corresponding virtual environment. However, unlike general man-made objects, trees usually have more complex morphological structures . There are diverse morphological differences both among and within trees, which bring great challenges to realistic modeling and rendering.

Over the past few decades, researchers from different backgrounds have been working to combine knowledge from multiple disciplines to address multiple types of challenges. Many valuable scientific questions and new methods have been continuously discovered and proposed, thereby enabling trees Modeling eventually developed into an interdisciplinary research involving multiple disciplines, among which natural sciences, mathematics and computer science have contributed to the development of the entire field of tree modeling.

Over the past few decades, researchers from different backgrounds have been working to combine knowledge from multiple disciplines to tackle multiple types of challenges. Many valuable scientific questions and new methods have been continuously discovered and proposed, thereby enabling trees Modeling eventually developed into an interdisciplinary research involving multiple disciplines, in which natural science, mathematics and computer science played a leading role in the development of the entire field of tree modeling. It is precisely because these disciplines actively promote each other , so that the research on tree modeling has been constantly moving forward. In verifying the corresponding hypothesis in botany, computer simulation provides a very effective means; at the same time, some new discoveries and corresponding theories in botany The development of tree modeling will provide a more purposeful direction of improvement for computer simulation. Based on these characteristics, when discussing the work in tree modeling, it can be compared and analyzed from the standpoint of different disciplines. This paper from the perspective of computer graphics Starting out, the related work of tree modeling is organized and discussed.

In computer graphics, modeling and rendering are two core research problems. For this reason, this paper will discuss the related work of trees from these two aspects. It should be pointed out that the work of these two parts is not independent, Rather, there are correlations. For example, many works in vegetation modeling have designed algorithms with efficient photorealistic rendering as a primary goal. This paper focuses on the photorealistic morphological structure of individual trees, leaves and their bark Modeling and rendering. Since these are the basic units that constitute complex scenes, and many of the research problems and corresponding solutions have great commonalities, the analysis from these basic units can simplify the corresponding discussions, highlighting The most important problems and corresponding solutions.

For the application of computer graphics, the current research on tree modeling focuses on the realistic modeling and presentation of the shape and appearance of trees. For this reason, the core problem to be solved is how to efficiently model and draw certain plants. Learn photorealistic tree models. When studying a single tree, the main objects to be modeled are the tree's branch structure, leaves, and bark. During tree growth, many of the internal generation that affects its structure is currently not fully understood. However, related studies such as plant morphogenesis still provide a corresponding theoretical basis for computer simulation of natural tree growth. According to the different principles and inputs on which tree branch modeling is based, this paper divides these works into four categories for analysis and discussion. These four categories include: rule-based procedural modeling, reconstruction based on video images and point clouds , sketch-based modeling and inverse modeling.

For leaves and bark, unless there is a special requirement for the realism of the model, the effect can generally be easily achieved by techniques such as texture mapping in graphics. Considering some other actual application requirements, this paper analyzes this part of the work , will mainly discuss methods involving accurate geometric modeling and photorealistic rendering. Based on the key problems in these methods, this paper will focus on solving these problems. The latest progress made is analyzed and summarized. The overall mapping of trees relies on special processing of tree branch structure to support efficient computation, so this part of the work will combine the discussion of tree branch modeling work in the above four categories at the same time. The goal of this paper is to systematically sort out the key issues and core technologies in the process of photorealistic modeling and rendering of trees through the analysis and discussion of the above work, so as to provide valuable exploration for future research in this field.

\section{Modeling and Rendering of Trunks and Branches}

According to the means of data acquisition and the difference in the form of data, the work on tree branch modeling can be divided into two categories: forward modeling and reverse modeling. Some simple rules are used to describe the branching characteristics of trees; by applying these rules repeatedly, a tree model with a certain sense of reality can finally be generated. However, no matter in the diversity of morphology or the complexity of branch structure Tree models are far from real-world trees. Physically, from the root to the branches, the bottom-up modeling follows the tree's botanical growth rules, which are collectively referred to as forward modeling in this paper. Methods. For this type of method, when the formation mechanism of corresponding characteristics in botany has a deeper understanding, more abundant rules can be defined and used to simulate the growth process of trees, and more realistic tree models can be generated.

With the popularization of data acquisition technologies such as video images, people can measure the modeling objects more conveniently, which provides a lot of data for related research on tree modeling. Based on these data, the research on tree modeling can not only be limited In addition to the method in forward modeling, it can also be modeled in reverse. Therefore, these video images or 3D point cloud data obtained by laser scanning not only open up new research directions, but also bring new challenges. In this paper, this kind of method that starts from the overall tree data and deduces its branch structure or generative model from the global to the local way is collectively referred to as the reverse modeling method. The original signal is accurately recovered from the sampled data. Since the inverse problem is essentially a kind of ill-conditioned problem, the existence or uniqueness of its solution cannot be guaranteed. Therefore, in practical applications, it is usually necessary to impose certain constraints or introduce corresponding prior knowledge In order to obtain a solution that meets specific needs. According to the difference of the input and the content to be deduced, this paper further subdivides such methods into three categories: reconstruction based on video images and point clouds, sketch-based modeling, and other reverse reconstruction methods. model method.

For the work on forward modeling of trees, this paper mainly introduces the rule-based procedural modeling method. The development of such methods has gone through the early structured modeling and the current cutting-edge hybrid method that integrates structured and self-organized modeling. Modeling. For inverse modeling, this paper first discusses these modeling methods based on video images and sketches, and then analyzes some other more challenging inverse modeling methods.

\section{Rule-based Procedural Modeling Approach}

\subsection{Structured Modeling Based on Geometric Rules}

For rule-based procedural modeling methods, the rules used generally describe the simple growth characteristics of branches and the corresponding geometric information. In earlier works, the geometric rules used in modeling are generally relatively simple. References \cite{r1} Two types of geometric information, namely branch angle and branch length ratio, are used to describe the branch structure of the tree. By adjusting the branch angle, the width and stretching degree of the overall shape of the tree can be controlled; The relative proportion of branch lengths can control the apical dominance of the tree to a certain extent, which will affect the overall outline of the tree. The method in \cite{r2} is also based on geometric rules, but it uses more abundant rules to support A variety of geometric generative models; finally, by combining these geometric models, more abundant branching characteristics can be simulated. Literature \cite{r3,r4,r5,r6} further introduced new rules based on the previous work based on botanical rules and measured data The new botanical factors considered are the mechanism of maximizing effective leaf area satisfied by the distribution of tree branches, and the modulation of competition among tree branches. The tree model generated by these methods has a certain sense of reality because of its branch and trunk structure to a certain extent, which is consistent with the actual observation data. Tree species, there are certain limitations in modeling tree models of other tree species.

\begin{figure}
    \centering
    \includegraphics[width=0.48\textwidth]{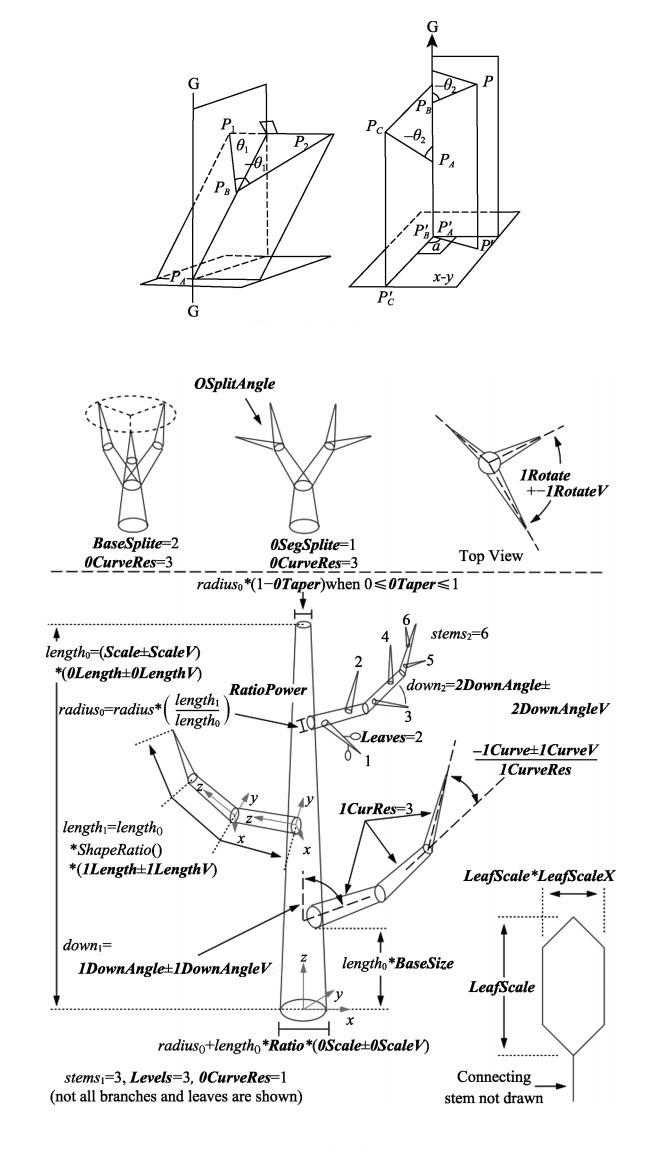}
    \caption{Schematic diagram of the geometric parameters used by the method in the literature \protect\cite{r1, r7} (top and bottom).}
    \label{fig1}
\end{figure}

Considering the limitation of simple geometric rules in the methods such as literature \cite{r1,r2} on the realism of the model, literature \cite{r7} introduced a large number of geometric parameters to finely control the branching structure of trees at all levels, since this method focuses more on controlling The specific geometry of the tree branches relaxes the botanical principles that the branches need to follow. Finally, the model proposed by this method uses a total of about 80 parameters to control the tree branch structure; in the actual modeling, In order to model a specific tree, a large number of parameters need to be adjusted, and the modeling efficiency is not high. Figure \ref{fig1} shows the schematic diagram of the geometric control parameters of tree branches by the method of literature \cite{r1,r7}. For the detailed introduction of each geometric parameter, please refer to corresponding literature.

\subsection{Structured Modeling Based on Growth Rules}
In addition to using geometric parameters to directly control and generate the overall branches of trees, other works simulate the growth of trees by further analyzing the growth rules behind different morphological structures. Reference \cite{r8} regards the growth process of tree branches as a The iterative process of moving a basic geometry along a certain trajectory, the basic geometry used is a polyhedron similar to a prism, and the size and orientation of its bottom and top surfaces can be controlled by corresponding parameters; according to the preset growth rules, in When gathering these polyhedrons from the bottom to the top to simulate the corresponding growth process, it is necessary to continuously adjust their parameters to generate a realistic branch structure. However, the branches generated in this way are not smooth enough, and the growth rules used are too simple. , which will cause the final generated tree model to be unrealistic. For this reason, the literature \cite{r9} proposes a generalized cylinder technique to generate branches with smooth surfaces. First, the skeleton structure of the tree branches is generated, and then the surface is scanned by scanning the surface. Based on the skeleton structure, the final branch surface is generated; among them, the skeleton structure of the tree corresponds to a series of three-dimensional points and the connection between them. These three-dimensional points mainly correspond to the nodes of the branches, and their spatial positions can be Calculated using the above growth model.

\begin{figure}
    \centering
    \includegraphics[width=0.48\textwidth]{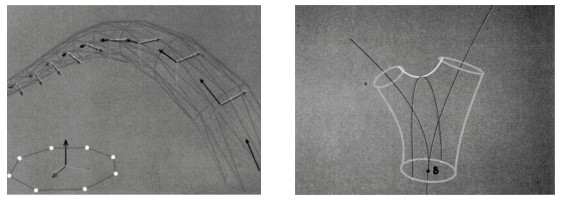}
    \caption{Surface modeling of branches and trunks based on generalized cylinders. Left: control of the local frame. Right: modeling of bifurcated branch trunk surfaces.}
    \label{fig2}
\end{figure}

\begin{figure}
    \centering
    \includegraphics[width=0.48\textwidth]{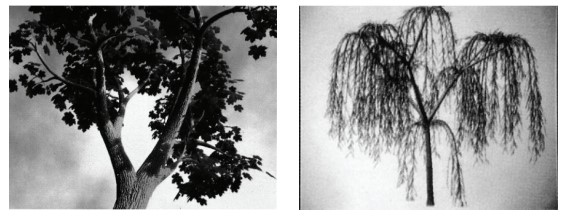}
    \caption{Methods for modeling trees \protect\cite{r9,r10}.}
    \label{fig3}
\end{figure}

When constructing the surface of the branch, first use the spline function to interpolate a series of new nodes inside the branch; then move a disk along these nodes by a certain step to build the corresponding cylindrical surface. In this Two points should be paid attention to in the process: (1) In order to avoid the distortion of the branch surface, the selection of the step size should ensure that the local frame at each interpolation point has only a smaller rotation than the previous position (as shown in \ref{fig2} left); (2) Special treatment should be performed on the branches of the branches to avoid the interpenetration between the surfaces of adjacent branches and to ensure the smoothness of their connections (as shown in Figure \ref{fig2} right). Because this method is relatively simple and can generate a smooth branch surface (as shown in Figure \ref{fig3} left), many current tree modeling works use this method to generate the branch surface.

Different from the modeling of the branch surface in Reference \cite{r9}, Reference \cite{r10} more faithfully follows the laws of botany to simulate the growth process of the whole tree. According to the corresponding principles in botany, Reference \cite{r10} The meristem in the tree bud is regarded as the growth engine of the whole tree, and its different "fate" during the growth process determines the morphological structure of the tree. For this reason, this method determines the different growth of the tree according to a certain probability distribution. stage, the three possible states of its meristem: growth, bifurcation and death. Similarly, for the newly grown branches, the parameters such as length, thickness and branch angle are also set through the corresponding probability distribution For different tree species, different probability distributions can be selected to control the above process. Since the model can combine external environmental influences (such as gravity, light, etc.) with the growth process, the generated model can be In terms of the richness of morphological structure and the botanical realism of the distribution of branches and trunks, a certain realistic effect can be achieved as shown in Figure \ref{fig3} (\cite{r10}).

The branching structure of trees is a strong evidence for the existence of fractal structures in nature \cite{r11}, and its self-similarity allows the creation of tree branch structures in a recursive manner. Reference \cite{r12} also used the The topological structure and its geometry are modeled separately. However, different from methods such as literature \cite{r9}, literature \cite{r12} builds a tree-like data structure in a recursive way based on the characteristics of the fractal structure. In the tree-like data structure, each A node represents a branch and stores a variety of geometric information, and the connection relationship between different nodes corresponds to the branches of the tree. Reference \cite{r12} does not use only a single geometry to represent branches as in Reference \cite{r8} , but defines 4 basic geometric types to model richer branch structures. The most basic of these geometric bodies is still a cylinder, and the rest are based on this further considering specific rotations, or introducing a certain randomness. The obtained spiral-like shape. It should be pointed out that in the literature \cite{r12}, all parameters describing the branch geometry are regarded as a kind of gene expression of a specific tree species. New tree models with different geometries. This feature that can easily support the generation of a large number of new tree models with different shapes is an important factor to consider when designing their tree modeling methods in many subsequent works.

\begin{figure}
    \centering
    \includegraphics[width=0.48\textwidth]{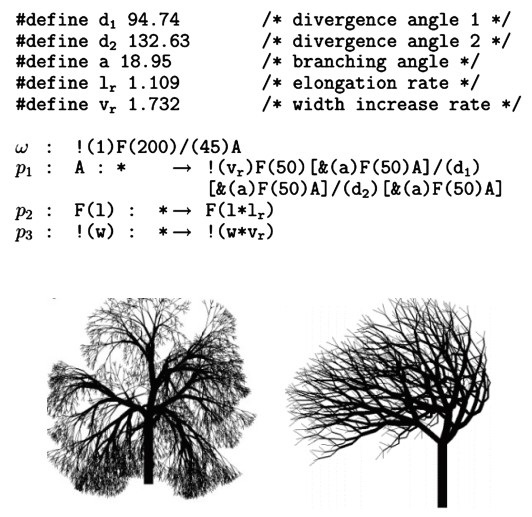}
    \caption{L-system and the corresponding tree model.}
    \label{fig4}
\end{figure}

For plant modeling, the L-system provides a modeling language for describing plant growth \cite{r13,r14}. Similar to the literature \cite{r12}, the final output string of the L-system can also be regarded as An expression of the corresponding branch structure \cite{r15}. In simple terms, an L-system is a string-based replacement and rewriting system consisting of an initial string called an axiom and a set of Substitution rules are composed. Generally, the L-system needs to distinguish 2 types of characters: terminators and non-terminators. Terminators generally have specific graphic meanings, and only non-terminators are allowed to be replaced. Starting from the axioms, each time by selecting the corresponding replacement rules of L-system rewrite non-terminal characters to generate new strings. In order to generate diverse models, L-system can not only specify a set of control parameters for characters, but also introduce certain parameters in the above replacement process. Condition or randomness. From a botanical point of view, different characters can be given corresponding botanical meanings, for example, characters can be used to represent plant organs such as tree buds or branches, so that the above-mentioned string represented by a string can be used. Substitution rules are regarded as the encoding of the corresponding growth rules in botany. For example, the botanical rules to be followed when branching out multiple lateral branches from a branch can be defined. Through this correspondence, it is not only convenient to write L-system grammars , which can also help others to understand. In this way, a parallel replacement process in the L-system can be regarded as a simulation of plant growth in a certain growth period, and the output string after repeating the process for a certain number of times, then The final structure of the plant is encoded. Based on this string, it also needs to be graphically interpreted accordingly to output the final geometric model. At present, this process is mainly completed by the turtle graphics system \cite{r13,r14}. Figure \ref{fig4} shows It is shown as an L-system grammar in \cite{r14}, and its two different tree models generated by the application of the turtle graphics system.

As can be seen from Figure \ref{fig4}, the L-system has a strong model expression ability. Because of this, the L-system is widely used in the field of tree modeling. However, for ordinary users, the L-system is directly applied Modeling trees is not an easy task. The reasons are as follows: (1) L-system grammars need to be written specifically for the tree species being modeled, and this requires certain expertise. (2) L-systems simulate In the forward growth process of plants, although the local characteristics of plants can be easily controlled, there is no direct control method for the overall characteristics. Therefore, this bottom-up method has certain limitations. (3) The L-system in \cite{r13,r14} lacks the interaction mechanism with the environment and cannot effectively simulate the changes in plant morphology due to the influence of the environment on the growth process. Therefore, many subsequent studies have tried to solve the L-system from different aspects. -These challenges of the system. Although some good progress has been made, there is still a lot of room for further exploration.

\begin{figure}
    \centering
    \includegraphics[width=0.48\textwidth]{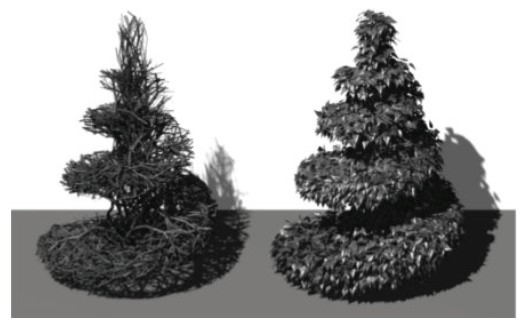}
    \caption{Modeling results based on overall shape.}
    \label{fig5}
\end{figure}

Reference \cite{r16} began to consider extending the L-system to support the corresponding role of the environment, so that the generation of the tree model can be controlled to a certain extent as a whole. The main extension is to make the L-system support querying the corresponding branches from the environment. The spatial position of the trunk is determined, and the pruning operation is performed on the branches beyond the corresponding spatial range. Specifically, once the spatial position of some branches reaches the boundary of the specified space, their subsequent growth will be restricted. The branches will eventually stop growing; and the resources it has gathered will be passed down to the unrestricted branches according to the feedback mechanism in botany, thereby promoting their growth. In this way, it is possible to Control the overall shape of the plant model. Figure \ref{fig5} shows the tree model under specific shape control illustrated in \cite{r16}.

The context-aware mechanism in \cite{r16} is further extended in \cite{r17} to support richer environmental interactions, and the system is called an open L-system. The L-system is a simulation system that supports the bidirectional interaction between plants and the environment. The interaction between plants and the environment considered in Reference \cite{r17} is mainly It includes three types of competition mechanisms: space competition, root competition for nutrients and water, and competition for sunlight. In order to further support the communication between plants and the environment, the literature \cite{r17} generalized the query of branches to the environment in literature \cite{r16}. The position command enables it to support the transfer of more parameters, thereby facilitating the data transfer during the communication between plants and the environment.

\begin{figure}
    \centering
    \includegraphics[width=0.48\textwidth]{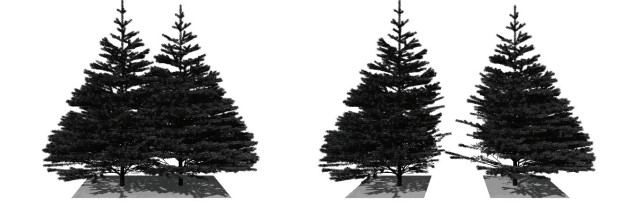}
    \caption{Effects of sunlight competition on tree branches.}
    \label{fig7}
\end{figure}

In the open L-system, when the plant receives the information from the environment, the corresponding functional modules in the L-system will respond. In this way, the growth process of the plant will adjust accordingly with these responses and output certain feedback; these feedbacks can then be queried by the environment through the same interface, creating a loop of information flow between the two. Figure \ref{fig7} shows the two pine tree models modeled in \cite{r17}, and we can see the effect of sunlight competition on the structure of tree branches. By simulating this sunlight competition mechanism, the distribution of branches in the final tree model will be No longer rules, and this is a realistic simulation of real situations in nature.

\subsection{Self-Organizing Tree Modeling}

It can be seen from the above analysis that methods such as literature \cite{r1,r8,r12,r13,r14,r16,r17} are all based on certain rules and use a recursive way to generate the final branch structure. This kind of method is generally used in the field of tree modeling. It is called a structured modeling method, and the model obtained by this modeling method is called a structured tree model, which is mainly suitable for simulating the early growth process of trees. When simulating mature trees, also it is necessary to further consider the spatial competition mechanism between branches and trunks in the process of tree growth \cite{r18}. Therefore, Reference \cite{r19} proposes a space colony algorithm to simulate the local space competition mechanism between branches and trunks in the process of tree growth. The principle of the algorithm is simple and easy to implement, and the generated tree model can ensure a certain botanical realism.

\begin{figure}
    \centering
    \includegraphics[width=0.48\textwidth]{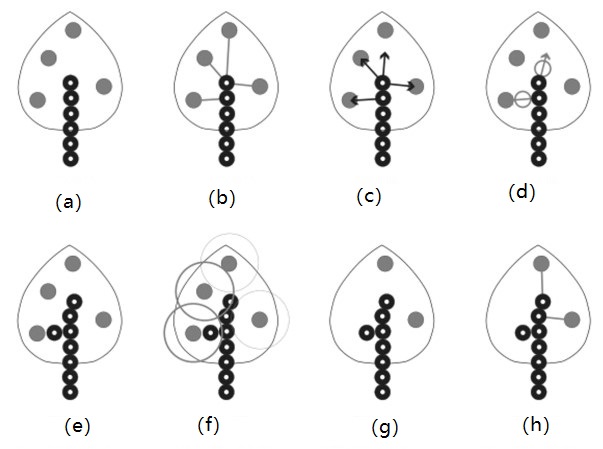}
    \caption{Flowchart of Space Colony Algorithm.}
    \label{fig8}
\end{figure}

On the whole, the process of generating tree branches in the literature \cite{r19} can be summarized into three steps: (1) It needs to randomly sprinkle points in a given spatial shape to characterize the space where the branches are allowed to grow; (2) Based on these spatial points, the growth of trees is simulated bottom-up to obtain the skeleton structure of the branches, which corresponds to the main process of the space colony algorithm; (3) Based on the skeleton structure, the surface geometry of the branches is generated, and the corresponding leaves are added by adding the corresponding leaves. etc. to enhance the realism of the final tree model. Figure \ref{fig8} shows the main flow of the spatial colony algorithm in the literature \cite{r19}.

\begin{figure}
    \centering
    \includegraphics[width=0.48\textwidth]{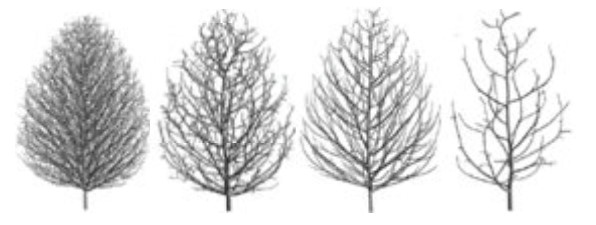}
    \caption{Tree models based on modeling with different parameters.}
    \label{fig9}
\end{figure}

In Figure \ref{fig8}a, the solid circles represent the randomly scattered spatial points at the beginning, and the hollow circles represent those nodes that constitute the current branch. In this way, the spatial positions of the branch that are allowed to grow in the next step are marked by these solid circles. At the beginning of the next iteration, for each node of the current branch, all the spatial points that fall within a certain radius neighborhood (as shown in  Figure \ref{fig8}b) are calculated, and the spatial points in these neighborhoods will determine the growth of new branches in the next step . The rules for generating new branches are used to determine the growth direction and the length of the branches (as shown in  Figure \ref{fig8}c). In  Figure \ref{fig8}d, the open circles represent the positions of new tree nodes determined by the specified length along the growth direction. After adding new tree nodes at these positions (as shown in  Figure \ref{fig8}e), it is also necessary to judge each spatial point in its neighborhood to decide whether to eliminate the spatial point. To this end, the algorithm for each spatial point A test radius is pre-specified, and if there is a tree node within the radius (as shown in  Figure \ref{fig8}f), the spatial point will be removed (as shown in  Figure \ref{fig8}g). In the next iteration, the above-mentioned steps are repeated according to the remaining spatial points. The process (shown in  Figure \ref{fig8}h). Figure \ref{fig9} shows the different tree models based on the same contour given in the literature [19] by setting different parameters in the above process.

The method of literature \cite{r19} can generate a variety of tree models, but the spatial colony algorithm only considers the local space competition mechanism of the branches and does not consider other important botanical factors (such as phyllotaxy, etc.) Both the local geometry and overall distribution of stems play a very important role, which will greatly enhance the botanical realism of tree models. These botanical factors cannot be fully simulated by structured modeling methods. For this reason, the literature \cite{r20} On the basis of this type of model, a self-organized tree modeling method is introduced. This self-organized modeling method is similar to the method in literature \cite{r10}, that is, it simulates the different fates of tree buds during the growth process. Control the generation of branches. However, the literature \cite{r10} mainly adopts a random way to control the growth of the branches, and the growth of the branches in the literature \cite{r19} also considers the external regulation mechanism of space and sunlight competition in the literature \cite{r19}, and the resource allocation and regulation mechanism inside the branch in the literature \cite{r5}.

The core modeling process based on the self-organization approach in literature \cite{r20} can be summarized into five steps:

\begin{enumerate}
    \item Calculate the local growth space and illumination of each tree bud (can be based on the spatial colony algorithm in \cite{r19} or based on the shadow propagation algorithm in \cite{r21}).
    \item The growth resources obtained by each tree bud, according to the similar text
    \item Adjust and redistribute using the internal allocation mechanism in \cite{r5}.
    \item Generate new branches based on the growth resources of each bud.
    \item Perform pruning test on each branch.
    \item Update the radius of all branches based on the current branch structure.
\end{enumerate}

\begin{figure}
    \centering
    \includegraphics[width=0.48\textwidth]{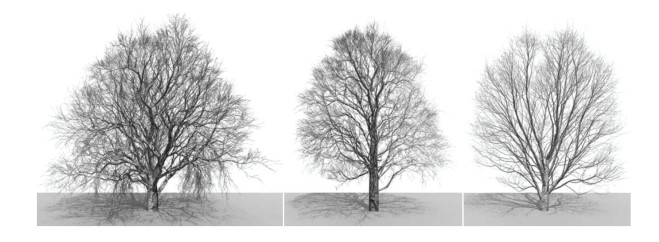}
    \caption{Different tree models obtained by controlling for apical dominance during tree growth.}
    \label{fig10}
\end{figure}

\begin{figure}
    \centering
    \includegraphics[width=0.48\textwidth]{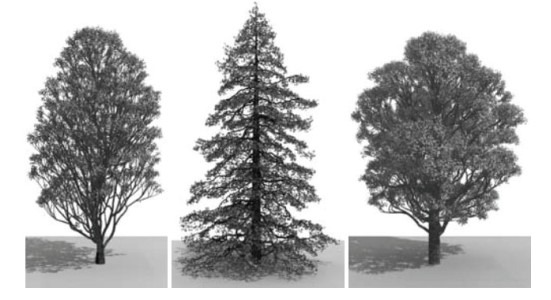}
    \caption{The three tree models further illustrated in \protect\cite{r20}.}
    \label{fig11}
\end{figure}

During the whole growth simulation process, these steps will be repeated to get the final ideal tree model. One of the advantages of the method in \cite{r20} is that it can effectively simulate some of the trees in the growth process by adjusting fewer parameters. Important botanical characteristics, such as apical dominance, etc., play a very important role in the formation of different morphological structures of trees. In addition, literature \cite{r20} combines two tree modeling methods, structured and self-organized, when adjusting some structural parameters. After the model, the branch density and overall distribution of the model will be automatically adjusted accordingly to ensure the naturalness and fidelity of the model. Figure \ref{fig10} shows the results obtained by controlling the apical dominance of different branches in literature \cite{r20}. 3 tree models. Figure \ref{fig11} further illustrates the tree model generated by the hybrid model in \cite{r20}. It can be seen that although the branches are subject to structural pruning and tropism (such as gravity) However, the final branch density and its spatial distribution maintain a more natural and realistic effect due to self-organization.

\subsection{Procedural Modeling Based on Hierarchical Graph Structure}

\begin{figure*}
    \centering
    \includegraphics[width=0.98\textwidth]{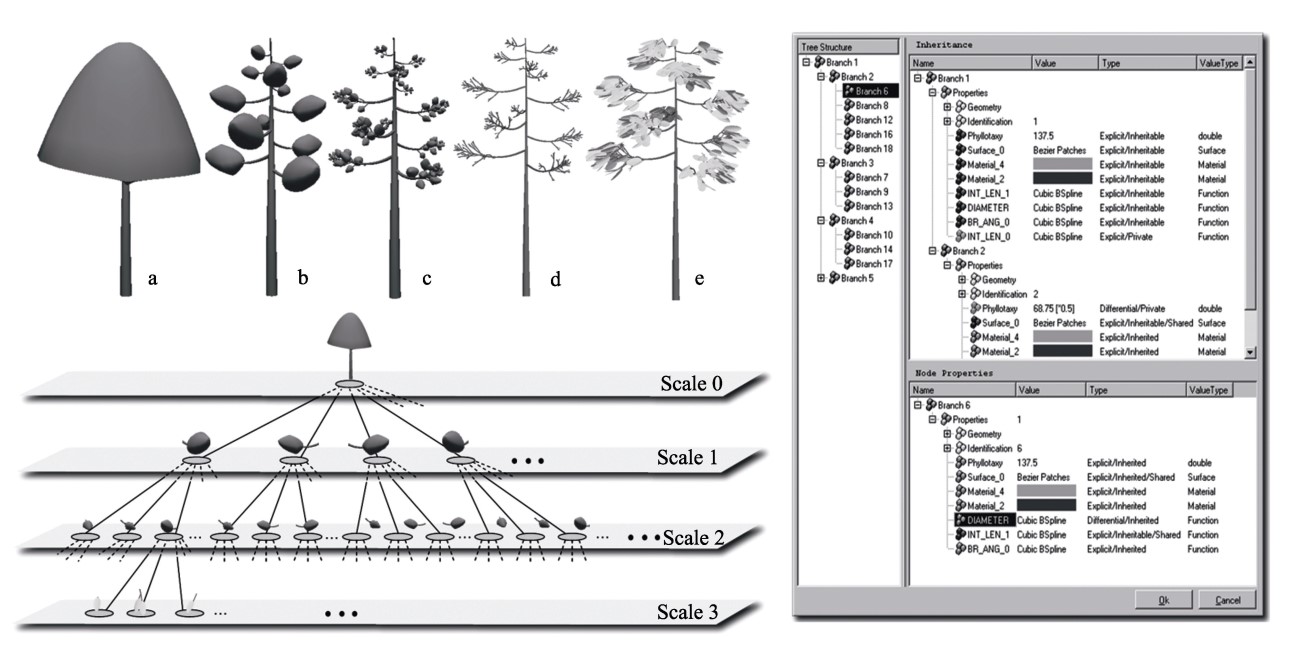}
    \caption{Branch-level tree diagram and its corresponding graphical parameter editor.}
    \label{fig12}
\end{figure*}

The above-mentioned bottom-up modeling methods are all modeled by simulating the growth rules of trees in nature; but similar to the L-system modeling method, the description of the local rules and the final tree morphological structure are closely related. There is no direct correlation, that is, it is impossible to directly judge the impact of changes in the corresponding parameters of local rules on the overall structure, and this is the source of its limitations in global control. For this reason, some methods combine a global-to-local modeling method, which not only provides control over the overall structure of trees, but also facilitates the adjustment of parameters in local rules. Reference \cite{r22} describes the overall posture of plants by introducing a graphical control method into the L-system. The spatial arrangement of L-system and its organs not only facilitates the user's global control over the growth process of the model, but also enhances the realism of the final model. Therefore, by designing a user-friendly graphical interface, it is possible to provide a more convenient L-system. The parameter adjustment method can alleviate the complexity of use to a certain extent. Based on the same idea, the literature \cite{r23} organizes the branches of trees into a tree-like hierarchical graph, and then provides a graph for the branches of each layer. In the tree hierarchy diagram, the top-level node corresponds to the trunk of the entire tree and its overall outline, and the second level corresponds to the side branches directly connected to the trunk. and its shape. In this way, by recursively applying similar decomposition rules to each side branch, the tree-like decomposition diagram as shown in Figure \ref{fig12} and the corresponding visual parameter editing interface can be obtained.

\begin{figure}
    \centering
    \includegraphics[width=0.48\textwidth]{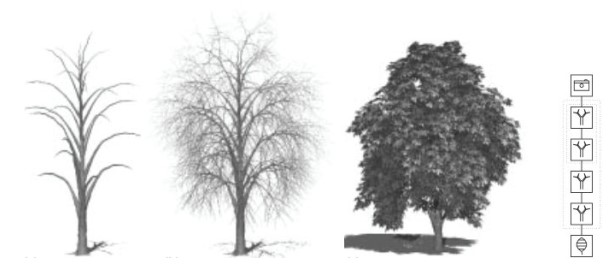}
    \caption{The p-graph corresponding to a single tree and its three-dimensional branch structure.}
    \label{fig13}
\end{figure}

By applying the idea of layered construction of tree branches, the complexity of the modeling process can be shared among different layers, simplifying the entire modeling process. Reference \cite{r24}pointed out that the main difficulty in tree modeling lies in Deal with the overall structure of the tree and its geometric complexity. In this way, a tree can be decomposed into different parts, and each part can be guaranteed to have an appropriate complexity to facilitate the user's modeling operations. Interaction in \cite{r24} A plant modeling system based on this idea is designed based on this idea, but it can support the modeling of a wider range of plant types, not just limited to trees. A set of components with different functions is defined in \cite{r24}, by grouping these components according to A directed graph (called p-graph) can be combined to describe the entire tree model; among them, it contains a class of basic components for creating basic plant organs, and other components provide functions such as repetition of basic components and global control. Similar to literature \cite{r23}, a unified graphical interface is designed in literature \cite{r24} for intuitive operation of p-graph and the components it contains. Applying this modeling system, Figure \ref{fig13} shows the literature An example of modeling a tree in \cite{r24}, the tree model is mainly composed of 4 layers of branches and 1 layer of leaves. Generally speaking, Reference \cite{r24} is also a procedural modeling method, and its components also use \cite{r9,r12} used a method to control and construct the geometry of branches; however, \cite{r24} provided a more concise and intuitive way to control the entire modeling process both globally and locally.

\begin{figure}
    \centering
    \includegraphics[width=0.48\textwidth]{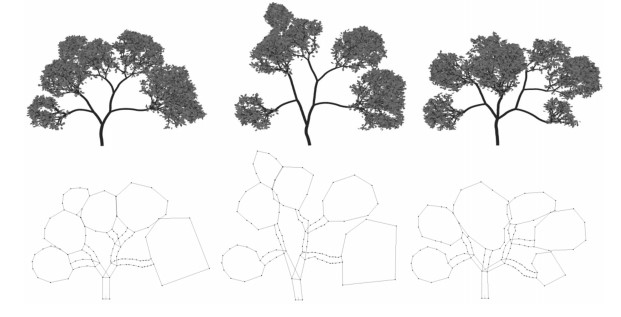}
    \caption{Tree modeling effect based on substructure.}
    \label{fig14}
\end{figure}

Literature \cite{r19,r20,r22,r23,r24} and other methods provide a variety of ideas to solve the limitations of the bottom-up tree generation method. Literature \cite{r25} decomposes the idea of tree structure in literature \cite{r23,r24} Combined with the communication mechanism between plants and the environment in \cite{r17}, a new top-down tree modeling method is proposed. In \cite{r25}, the whole tree is decomposed into different substructures, and each substructure can be Control is carried out by a procedural method similar to that in the L-system. In this way, compared with the whole tree, each substructure is not only easy to control, but also can be conveniently organized in a certain global way to achieve the overall control effect. Figure \ref{fig14} Shown is the modeling effect of document \cite{r25}controlling the tree model from global to local, which shows the corresponding substructure of each tree and the change of the overall structure of the tree after corresponding adjustment. Reference \cite{r25}, which is widely used by methods such as inverse modeling to be discussed later. For inverse modeling work, it usually requires top-down modeling based on the input global information. model the corresponding branch structure.

\section{Conclusion}

At present, rule-based procedural modeling work mainly reflects two development trends: (1) The transition from structured modeling to self-organizing hybrid modeling. This kind of hybrid modeling method combines growth rules and environmental effects (2) From the non-intuitive modeling control methods such as parameters and grammar rules to the procedural modeling method based on intuitive graphical structure Conversion. The improvement of this interactive modeling method greatly improves the modeling efficiency of users, so it is adopted by today's mainstream commercial tree modeling software.

Additionally, there are some extended articles about applying physical lighting computation in various applications:

\begin{enumerate}
    \item Deep Learning-Based Monte Carlo Noise Reduction By training a neural network denoiser through offline learning, it can filter noisy Monte Carlo rendering results into high-quality smooth output, greatly improving physics-based Availability of rendering techniques \cite{huo2021survey}, common research includes predicting a filtering kernel based on g-buffer \cite{bako2017kernel}, using GAN to generate more realistic filtering results \cite{xu2019adversarial}, and analyzing path space features Perform manifold contrastive learning to enhance the rendering effect of reflections \cite{cho2021weakly}, use weight sharing to quickly predict the rendering kernel to speed up reconstruction \cite{fan2021real}, filter and reconstruct high-dimensional incident radiation fields for unbiased reconstruction rendering guide \cite{huo2020adaptive}, etc.
    \item The multi-light rendering framework is an important rendering framework outside the path tracing algorithm. Its basic idea is to simplify the simulation of the complete light path illumination transmission after multiple refraction and reflection to calculate the direct illumination from many virtual light sources, and provide a unified Mathematical framework to speed up this operation \cite{dachsbacher2014scalable}, including how to efficiently process virtual point lights and geometric data in external memory \cite{wang2013gpu}, how to efficiently integrate virtual point lights using sparse matrices and compressed sensing \cite{huo2015matrix}, and how to handle virtual line light data in translucent media \cite{huo2016adaptive}, use spherical Gaussian virtual point lights to approximate indirect reflections on glossy surfaces \cite{huo2020spherical}, and more.
    \item Automatic optimization of rendering pipelines Apply high-quality rendering technology to real-time rendering applications by optimizing rendering pipelines. The research contents include automatic optimization based on quality and speed \cite{wang2014automatic}, automatic optimization for energy saving \cite{ wang2016real,zhang2021powernet}, LOD optimization for terrain data \cite{li2021multi}, automatic optimization and fitting of pipeline rendering signals \cite{li2020automatic}, anti-aliasing \cite{zhong2022morphological}, etc.
    \item Using physically-based process to guide the generation of data for single image reflection removal \cite{kim2020single}; propagating local image features in a hypergraph for image retreival \cite{an2021hypergraph}; managing 3D assets in a block chain-based distributed system \cite{park2021meshchain}.
\end{enumerate}


\bibliographystyle{ieee}
\bibliography{srbib}

\begin{thebibliography}{10}\itemsep=-1pt

\bibitem{an2021hypergraph}
G.~An, Y.~Huo, and S.-E. Yoon.
\newblock Hypergraph propagation and community selection for objects retrieval.
\newblock {\em Advances in Neural Information Processing Systems}, 34, 2021.

\bibitem{r2}
M.~Aono and T.~L. Kunii.
\newblock Botanical tree image generation.
\newblock {\em IEEE computer graphics and applications}, 4(5):10--34, 1984.

\bibitem{bako2017kernel}
S.~Bako, T.~Vogels, B.~McWilliams, M.~Meyer, J.~Nov{\'a}k, A.~Harvill, P.~Sen,
  T.~Derose, and F.~Rousselle.
\newblock Kernel-predicting convolutional networks for denoising monte carlo
  renderings.
\newblock {\em ACM Trans. Graph.}, 36(4):97--1, 2017.

\bibitem{r25}
B.~Bene{\v{s}}, O.~{\v{S}}t'ava, R.~M{\v{e}}ch, and G.~Miller.
\newblock Guided procedural modeling.
\newblock In {\em Computer graphics forum}, volume~30, pages 325--334. Wiley
  Online Library, 2011.

\bibitem{r9}
J.~Bloomenthal.
\newblock Modeling the mighty maple.
\newblock {\em ACM SIGGRAPH Computer Graphics}, 19(3):305--311, 1985.

\bibitem{r5}
R.~Borchert and H.~Honda.
\newblock Control of development in the bifurcating branch system of tabebuia
  rosea: a computer simulation.
\newblock {\em Botanical Gazette}, 145(2):184--195, 1984.

\bibitem{r21}
S.~Bornhofen and C.~Lattaud.
\newblock Competition and evolution in virtual plant communities: a new
  modeling approach.
\newblock {\em Natural Computing}, 8(2):349--385, 2009.

\bibitem{r23}
F.~Boudon, P.~Prusinkiewicz, P.~Federl, C.~Godin, and R.~Karwowski.
\newblock Interactive design of bonsai tree models.
\newblock In {\em Computer Graphics Forum}, volume~22, pages 591--599. Wiley
  Online Library, 2003.

\bibitem{cho2021weakly}
I.-Y. Cho, Y.~Huo, and S.-E. Yoon.
\newblock Weakly-supervised contrastive learning in path manifold for monte
  carlo image reconstruction.
\newblock {\em ACM Transactions on Graphics (TOG)}, 40(4):38--1, 2021.

\bibitem{dachsbacher2014scalable}
C.~Dachsbacher, J.~K{\v{r}}iv{\'a}nek, M.~Ha{\v{s}}an, A.~Arbree, B.~Walter,
  and J.~Nov{\'a}k.
\newblock Scalable realistic rendering with many-light methods.
\newblock In {\em Computer Graphics Forum}, volume~33, pages 88--104. Wiley
  Online Library, 2014.

\bibitem{r10}
P.~De~Reffye, C.~Edelin, J.~Fran{\c{c}}on, M.~Jaeger, and C.~Puech.
\newblock Plant models faithful to botanical structure and development.
\newblock {\em ACM Siggraph Computer Graphics}, 22(4):151--158, 1988.

\bibitem{fan2021real}
H.~Fan, R.~Wang, Y.~Huo, and H.~Bao.
\newblock Real-time monte carlo denoising with weight sharing kernel prediction
  network.
\newblock In {\em Computer Graphics Forum}, volume~40, pages 15--27. Wiley
  Online Library, 2021.

\bibitem{r3}
J.~B. Fisher and H.~Honda.
\newblock Computer simulation of branching pattern and geometry in terminalia
  (combretaceae), a tropical tree.
\newblock {\em Botanical Gazette}, 138(4):377--384, 1977.

\bibitem{r1}
H.~Honda.
\newblock Description of the form of trees by the parameters of the tree-like
  body: Effects of the branching angle and the branch length on the shape of
  the tree-like body.
\newblock {\em Journal of theoretical biology}, 31(2):331--338, 1971.

\bibitem{r6}
H.~Honda, H.~Hatta, and J.~B. Fisher.
\newblock Branch geometry in cornus kousa (cornaceae): computer simulations.
\newblock {\em American Journal of Botany}, 84(6):745--755, 1997.

\bibitem{r4}
H.~Honda, P.~Tomlinson, and J.~B. Fisher.
\newblock Computer simulation of branch interaction and regulation by unequal
  flow rates in botanical trees.
\newblock {\em American Journal of Botany}, 68(4):569--585, 1981.

\bibitem{huo2020spherical}
Y.~Huo, S.~Jin, T.~Liu, W.~Hua, R.~Wang, and H.~Bao.
\newblock Spherical gaussian-based lightcuts for glossy interreflections.
\newblock In {\em Computer Graphics Forum}, volume~39, pages 192--203. Wiley
  Online Library, 2020.

\bibitem{huo2016adaptive}
Y.~Huo, R.~Wang, T.~Hu, W.~Hua, and H.~Bao.
\newblock Adaptive matrix column sampling and completion for rendering
  participating media.
\newblock {\em ACM Transactions on Graphics (TOG)}, 35(6):1--11, 2016.

\bibitem{huo2015matrix}
Y.~Huo, R.~Wang, S.~Jin, X.~Liu, and H.~Bao.
\newblock A matrix sampling-and-recovery approach for many-lights rendering.
\newblock {\em ACM Transactions on Graphics (TOG)}, 34(6):1--12, 2015.

\bibitem{huo2020adaptive}
Y.~Huo, R.~Wang, R.~Zheng, H.~Xu, H.~Bao, and S.-E. Yoon.
\newblock Adaptive incident radiance field sampling and reconstruction using
  deep reinforcement learning.
\newblock {\em ACM Transactions on Graphics (TOG)}, 39(1):1--17, 2020.

\bibitem{huo2021survey}
Y.~Huo and S.-e. Yoon.
\newblock A survey on deep learning-based monte carlo denoising.
\newblock {\em Computational Visual Media}, 7(2):169--185, 2021.

\bibitem{r8}
Y.~Kawaguchi.
\newblock A morphological study of the form of nature.
\newblock In {\em Proceedings of the 9th annual conference on Computer graphics
  and interactive techniques}, pages 223--232, 1982.

\bibitem{kim2020single}
S.~Kim, Y.~Huo, and S.-E. Yoon.
\newblock Single image reflection removal with physically-based training
  images.
\newblock In {\em Proceedings of the IEEE/CVF Conference on Computer Vision and
  Pattern Recognition}, pages 5164--5173, 2020.

\bibitem{li2020automatic}
S.~Li, R.~Wang, Y.~Huo, W.~Zheng, W.~Hua, and H.~Bao.
\newblock Automatic band-limited approximation of shaders using mean-variance
  statistics in clamped domain.
\newblock In {\em Computer Graphics Forum}, volume~39, pages 181--192. Wiley
  Online Library, 2020.

\bibitem{li2021multi}
S.~Li, C.~Zheng, R.~Wang, Y.~Huo, W.~Zheng, H.~Lin, and H.~Bao.
\newblock Multi-resolution terrain rendering using summed-area tables.
\newblock {\em Computers \& Graphics}, 95:130--140, 2021.

\bibitem{r15}
A.~Lindenmayer.
\newblock Mathematical models for cellular interactions in development i.
  filaments with one-sided inputs.
\newblock {\em Journal of theoretical biology}, 18(3):280--299, 1968.

\bibitem{r24}
B.~Lintermann and O.~Deussen.
\newblock Interactive modeling of plants.
\newblock {\em IEEE Computer Graphics and Applications}, 19(1):56--65, 1999.

\bibitem{r11}
B.~B. Mandelbrot and B.~B. Mandelbrot.
\newblock {\em The fractal geometry of nature}, volume~1.
\newblock WH freeman New York, 1982.

\bibitem{r17}
R.~M{\v{e}}ch and P.~Prusinkiewicz.
\newblock Visual models of plants interacting with their environment.
\newblock In {\em Proceedings of the 23rd annual conference on Computer
  graphics and interactive techniques}, pages 397--410, 1996.

\bibitem{r12}
P.~E. Oppenheimer.
\newblock Real time design and animation of fractal plants and trees.
\newblock {\em ACM SiGGRAPH Computer Graphics}, 20(4):55--64, 1986.

\bibitem{r20}
W.~Palubicki, K.~Horel, S.~Longay, A.~Runions, B.~Lane, R.~M{\v{e}}ch, and
  P.~Prusinkiewicz.
\newblock Self-organizing tree models for image synthesis.
\newblock {\em ACM Transactions On Graphics (TOG)}, 28(3):1--10, 2009.

\bibitem{park2021meshchain}
H.~Park, Y.~Huo, and S.-E. Yoon.
\newblock Meshchain: Secure 3d model and intellectual property management
  powered by blockchain technology.
\newblock In {\em Computer Graphics International Conference}, pages 519--534.
  Springer, 2021.

\bibitem{r13}
P.~Prusinkiewicz.
\newblock Applications of l-systems to computer imagery.
\newblock In {\em International Workshop on Graph Grammars and Their
  Application to Computer Science}, pages 534--548. Springer, 1986.

\bibitem{r16}
P.~Prusinkiewicz, M.~James, and R.~M{\v{e}}ch.
\newblock Synthetic topiary.
\newblock In {\em Proceedings of the 21st annual conference on Computer
  graphics and interactive techniques}, pages 351--358, 1994.

\bibitem{r14}
P.~Prusinkiewicz and A.~Lindenmayer.
\newblock {\em The algorithmic beauty of plants}.
\newblock Springer Science \& Business Media, 2012.

\bibitem{r22}
P.~Prusinkiewicz, L.~M{\"u}ndermann, R.~Karwowski, and B.~Lane.
\newblock The use of positional information in the modeling of plants.
\newblock In {\em Proceedings of the 28th annual conference on Computer
  graphics and interactive techniques}, pages 289--300, 2001.

\bibitem{r19}
A.~Runions, B.~Lane, and P.~Prusinkiewicz.
\newblock Modeling trees with a space colonization algorithm.
\newblock {\em NPH}, 7:63--70, 2007.

\bibitem{r18}
T.~Sachs and A.~Novoplansky.
\newblock Tree form: architectural models do not suffice.
\newblock {\em Israel Journal of Plant Sciences}, 43(3):203--212, 1995.

\bibitem{wang2013gpu}
R.~Wang, Y.~Huo, Y.~Yuan, K.~Zhou, W.~Hua, and H.~Bao.
\newblock Gpu-based out-of-core many-lights rendering.
\newblock {\em ACM Transactions on Graphics (TOG)}, 32(6):1--10, 2013.

\bibitem{wang2014automatic}
R.~Wang, X.~Yang, Y.~Yuan, W.~Chen, K.~Bala, and H.~Bao.
\newblock Automatic shader simplification using surface signal approximation.
\newblock {\em ACM Transactions on Graphics (TOG)}, 33(6):1--11, 2014.

\bibitem{wang2016real}
R.~Wang, B.~Yu, J.~Marco, T.~Hu, D.~Gutierrez, and H.~Bao.
\newblock Real-time rendering on a power budget.
\newblock {\em ACM Transactions on Graphics (TOG)}, 35(4):1--11, 2016.

\bibitem{r7}
J.~Weber and J.~Penn.
\newblock Creation and rendering of realistic trees.
\newblock In {\em Proceedings of the 22nd annual conference on Computer
  graphics and interactive techniques}, pages 119--128, 1995.

\bibitem{xu2019adversarial}
B.~Xu, J.~Zhang, R.~Wang, K.~Xu, Y.-L. Yang, C.~Li, and R.~Tang.
\newblock Adversarial monte carlo denoising with conditioned auxiliary feature
  modulation.
\newblock {\em ACM Trans. Graph.}, 38(6):224--1, 2019.

\bibitem{zhang2021powernet}
Y.~Zhang, R.~Wang, Y.~Huo, W.~Hua, and H.~Bao.
\newblock Powernet: Learning-based real-time power-budget rendering.
\newblock {\em IEEE Transactions on Visualization and Computer Graphics}, 2021.

\bibitem{zhong2022morphological}
Y.~Zhong, Y.~Huo, and R.~Wang.
\newblock Morphological anti-aliasing method for boundary slope prediction.
\newblock {\em arXiv preprint arXiv:2203.03870}, 2022.

\end{thebibliography}

\end{document}